%% file: alpha.tex
\documentclass[aps,twocolumn,showpacs]{revtex4}
\usepackage[english]{babel}

\usepackage{graphicx}
\usepackage{dcolumn}
\usepackage{amsmath}

\def\beq#1{\begin{equation}\label{#1}}
\def\eeq{\end{equation}}

\begin{document}
\title{Phase properties of the cut-off high-order harmonics}
\author{ M.A.Khokhlova$^{1,2}$,
V.V. Strelkov$^{1,3}$
}
\affiliation{
$^1$A. M. Prokhorov General Physics Institute of the RAS, Moscow 119991,  Russia \\
$^2$Faculty of Physics, M.V. Lomonosov Moscow State University, Moscow 119991, Russia\\ 
$^3$Moscow Institute of Physics and Technology (State University), 141700 Dolgoprudny, Moscow Region, Russia 
}
\date{\today}
\begin{abstract}\noindent 
The cut-off regime of high-order harmonic generation (HHG) by atoms in an intense laser field is studied numerically and analytically. We find that the cut-off regime is characterized by equal dephasing between the successive harmonics. The change of the harmonic phase-locking when HHG evolves from the cut-off to the plateau regime determines the optimal bandwidth of the spectral region which should be used for attosecond pulse generation via amplitude gating technique. The cut-off regime is also characterized by a linear dependence of the harmonic phase on the fundamental intensity. The proportionality coefficient grows as the cube of the fundamental wavelength, thus this dependence becomes very important for the HHG by mid-infrared fields. Moreover, for every high harmonic there is a {\it range} of laser intensities providing the generation in the cut-off regime and the atomic response magnitude in this regime can be greater than that in the plateau regime. Thus the cut-off regime substantially contributes to the harmonic energy emitted under typical experimental conditions where the laser intensity varies in time and space.

\end{abstract}
\pacs{
  42.65.Ky            
  32.80.Wr	          
}
\maketitle
\noindent

\input{full.txt}

\input{appendix.txt}


\end{document}

%% file: full.txt
One of the most characteristic features of the high-order harmonic generation (HHG) by atoms, molecules and ions in intense laser field is the plateau in the spectrum, namely the region where the harmonic intensity almost does not depend on the harmonic order. The high-frequency part of the plateau ends up with a sharp decrease known as the cut-off of the harmonic spectrum~\cite{McPherson,Ferray,Li}. The number of harmonics in the plateau increases with the fundamental intensity~\cite{Li,Krause}. When the fundamental intensity increases the given harmonic first is generated at the cut-off and then within the plateau. At first glance, the former stage can be understood just as a sudden "turning on" of the generation, thus providing negligible contribution to the total signal. Using quantum mechanical calculations we show in this Letter that this is not the case. The cut-off regime can be attributed to a {\it range} of fundamental intensities and provides an important impact to the generation in typical experimental conditions. The dependence of the harmonic phase on the laser intensity substantially defines the phase-matching of the generation~\cite{Lew_Sal_lHuil, Peatross_phase-matching}, spectral shift and the harmonic line broadening~\cite{Peatross_shift,Platonenko_shift,nat_comm}, coherence~\cite{coherence} and the spatial properties of the harmonic beam~\cite{Lew_Sal_lHuil, Peatross_spatial, Gaarde_spatiotemporal, nat_comm}. Thus, the cut-off high harmonic phase behavior has a pronounced effect on these properties. Moreover, the cut-off xuv is used to generate an isolated attosecond pulse via amplitude gating technique~\cite{amp_gating_christov,amp_cep_plat,amp_cep_paulus,amp_cep_exp,amp_gate_exp}. Studying the phase properties of these harmonics allows us to suggest the spectral range which should be used in this method to provide the shortest attosecond pulse for given pump intensity and frequency.  

To analyze the HHG cut-off region theoretically we use a fully analytical result for the dipole moment~\cite{Becker_zrp} describing HHG emission by one electron bound to a zero-range potential in the monochromatic laser field.
The dipole moment of the $q^{th}$ harmonic ($q=2k+1$) generated in the linearly polarized laser field $E(t)=E_{0}\sin{\omega t}$ is presented in~\cite{Becker_zrp} as 

\begin{widetext}
\begin{center}
\begin{equation}
d_{2k+1}=\frac{8}{\omega}\left(\pi\frac{U_p}{\omega}I_p\right)^{1/2}
\frac{(-1)^k}{(2k+1)^2}\int_0^{\infty}\frac{d\tau}{\tau^{3/2}}
e^{i(k+1/2)\tau}(J_{k+1}(z(\tau))\beta_{+}(\tau)\sin{\alpha_{00k}(\tau)}-
J_{k}(z(\tau))\beta_{-}(\tau)\cos{\alpha_{00k}(\tau)})
\label{d_becker}
\end{equation}
\end{center}
\end{widetext}
where $I_p$ is the ionization potential of the generating particle, $U_p=\frac{E_0^2}{4\omega^2}$ is the ponderomotive energy, $\tau$ is the electron excursion time and functions $z(\tau)$, $\beta_{\pm}(\tau)$ and $\alpha_{00k}(\tau)$ are defined as follows
\begin{equation}
\begin{array}{l}
\beta_{\pm}(\tau)=\frac{e^{i\tau/2}}{1\pm\frac{1}{2k+1}}-
\frac{2}{\tau}\sin{(\tau /2)}\\
z(\tau)=\frac{U_p}{\omega}\left(\sin{\tau}-\frac{4\sin^2{(\tau/2)}}{\tau}\right)\\
\alpha_{00k}(\tau)=\frac{I_p}{\omega}\tau+\frac{U_p}{\omega}\tau\left(1-\left(\frac{\sin{(\tau/2)}}{\tau/2}\right)^2\right)
-\frac{\pi}{4}(2k+1)\\
\end{array}
\end{equation}

The harmonic phase dependence on the fundamental intensity was investigated in numerous studies: experimentally \cite{Yost}, numerically using the time-dependent Schr\"{o}dinger equation (TDSE) solution \cite{Gaarde_2002,Gaarde_1999}, theoretically within the SFA~\cite{Lew_Sal_lHuil,Balcou,Gaarde_2002,Antoine_1996}, using Feynman's path-integral approach \cite{Pascal_sci},  and using generalized semiclassical model \cite{Gaarde_semi}. As was discussed in these works the harmonic phase dependence can be described with the phase coefficient $\alpha_{2k+1}=-\frac{\partial \varphi_{2k+1}}{\partial I}$ where $\varphi_{2k+1}$ is the phase of the spectral component $d_{2k+1}$. Considering the derivative of the dipole moment with respect to the laser intensity the phase coefficient $\alpha_{2k+1}$ can be found as
\begin{equation}
\alpha_{2k+1}=-Im\left(\frac{1}{d_{2k+1}}\frac{\partial d_{2k+1}}{\partial I}\right)
\label{alpha_def}
\end{equation}
Substituting (\ref{d_becker}) in (\ref{alpha_def}) we derive an analytical equation for the phase coefficient $\alpha_{2k+1}$ in the form
\begin{widetext}
\begin{center}
\begin{equation}
\label{alphafull}
\begin{split}
\alpha_{2k+1}=\frac{1}{8\omega^3}
Im\left(
\frac{\int_0^{\infty}\frac{d\tau}{\tau^{3/2}}e^{i(k+\frac{1}{2})\tau}D(\tau)}
{\int_0^{\infty}\frac{d\tau}{\tau^{3/2}}e^{i(k+\frac{1}{2})\tau}
(J_{k+1}(z)\beta_{+}\sin{\alpha_{00k}}-J_{k}(z)\beta_{-}\cos{\alpha_{00k}})}\right){}\\
D(\tau)=\left(\sin{\tau}-\frac{4}{\tau}\sin^2{(\tau/2)}\right)\left((J_{k}(z)-J_{k+2}(z))\beta_{+}\sin{\alpha_{00k}}-
(J_{k-1}(z)-J_{k+1}(z))\beta_{-}\cos{\alpha_{00k}}\right){}\\
+2\tau\left(1-\left(\frac{\sin{(\tau/2)}}{\tau/2}\right)^2\right)
\left(J_{k+1}(z)\beta_{+}\cos{\alpha_{00k}}+J_{k}(z)\beta_{-}\sin{\alpha_{00k}}\right)
\end{split}
\end{equation}
\end{center}
\end{widetext}

\begin{figure}  
\centering
\includegraphics [width=0.9\columnwidth] {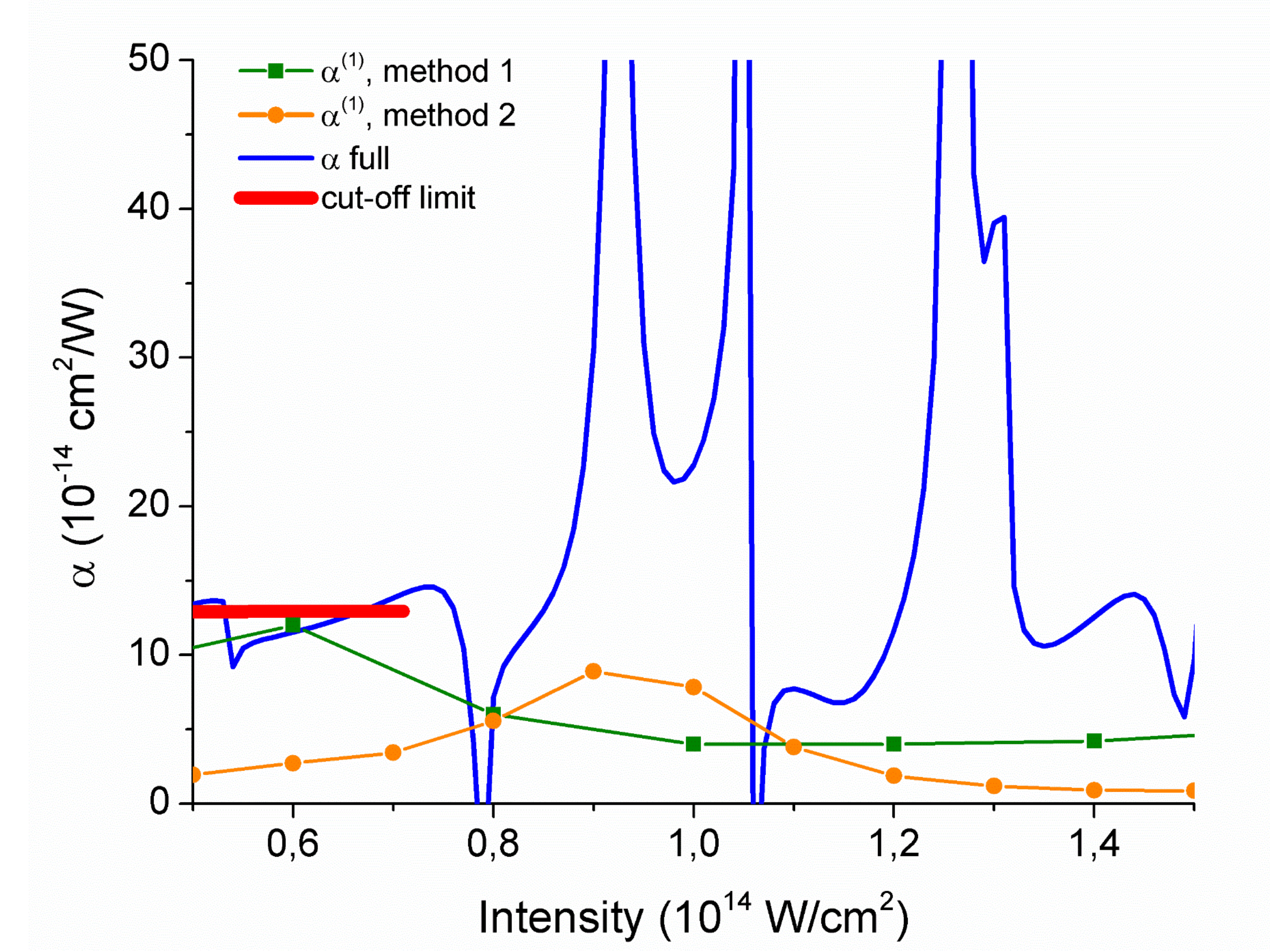}
\caption{(color online) The phase coefficient $\alpha$ for H17 generated in Xe as a function of laser intensity. The result obtained via equation~(\ref{alphafull}) (blue line) and the approximation for the cut-off region (\ref{alpha_cut-off}) (red thick line) are presented. Lines with symbols show the results of two methods based on the numerical 3D TDSE solution (the methods are described in the Appendix).}
\label{fig1}
\end{figure}

In Fig.~\ref{fig1} we present the phase coefficient $\alpha_{17}$ calculated  for H17 generated in Xe atoms by the laser field with the wavelength of 800 nm (below the frequency of this field is denoted as $\omega_{Ti:Sapp}$). One can see that there are two different regimes for the phase coefficient $\alpha$ behavior with respect to intensity. The first one (under lower intensities) corresponds to the cut-off region where the phase coefficient $\alpha$ is smooth, and the second one (under higher intensities) corresponds to the plateau region where the $\alpha$ behaves non-regularly. This can be explained by quantum paths interference~\cite{Amelle}. Thus, we define the cut-off region as the one where $\alpha$ almost doesn't depend on the laser intensity (according to a criterion that the value of $\alpha$ lays in between $\pm 20\%$  of its value for lowest intensity). The intensity corresponding to the change between these two regimes is denoted below as the cut-off/plateau transition intensity. Equation (\ref{alphafull}) for the phase coefficients can be simplified using the approximation for the end of the plateau and cut-off harmonics (see Eq.~(5.23) in \cite{Becker_zrp}). After some calculations one obtains as a limit for the cut-off region the following formula
 
\begin{equation}
\alpha \approx \frac{3.309}{4\omega^3}
\label{alpha_cut-off}
\end{equation}
Fig.~\ref{fig1} shows that this approximation agrees with the result of equation~(\ref{alphafull}) in the cut-off region. Moreover, in the same figure we present results of $\alpha$ calculation using two numerical methods discussed in detail in the Appendix. The methods are based on the numerical TDSE solution for a model one-electron atom with the technique described in~\cite{ref1}. The $\alpha$ values corresponding to the short trajectory only are presented in the plateau region (see lines with symbols in Fig.~ \ref{fig1}). We see that the method 1 reasonably agrees with the analytical results but for the method 2 the agreement is worth. When the field parameters are closer to the tunneling regime both numerical methods agree with the analytical result (see Figures 5 and 6 in the Appendix). The numerically found values for $\alpha$ are lower than the one for the long trajectory in the plateau regime but much higher than the one for the short trajectory. Thus, there is a significant difference between the cut-off and the plateau regime: in the latter there is at least the short trajectory contribution for which the phase dependence on the fundamental intensity can be neglected in many cases; for the cut-off regime this dependence is always considerable.  

In Fig.~\ref{fig2} the result for the cut-off/plateau transition intensity obtained by the criterion discussed above for the different fundamental frequencies is presented with red curve. The region below and above the red curve corresponds to the cut-off and the plateau region, respectively. The green curves show the fundamental intensity at which the generation of the given harmonic starts: the light green curve shows this intensity found via the simple-man approach~\cite{simple-man1,simple-man2} and the dark green one shows the result found using the exact cut-off law~\cite{Lew}. Below we denote the exact cut-off intensity as $I_{appear}$. One can see that the cut-off/plateau transition can be approximated as $1.3 I_{appear}$ (blue dashed curve) both for the case of H17 in Xe and H91 in He. 

The generation takes place in the cut-off regime for the range of fundamental intensities shown by the hatched area between the red and dark green curves in Fig.~\ref{fig2}. Within this intensity range the amplitude of the microscopic harmonic response achieves its maximum (see Figs.4-6 in Appendix) or it is close to the maximum. Thus, in the harmonic response experimentally generated for the laser intensity varying in time and in space the contribution of the cut-off regime is significant for many harmonics. Moreover, the laser intensity providing efficient HHG is practically limited by the medium ionization. Typically, several percent of the ionization provide the essential detuning from the phase-matching~\cite{phase-matching}.  This limitation practically cancels the HHG in the plateau regime for certain harmonics. To illustrate this we find numerically the intensities leading to more than $10\%$ ionization of Xe atoms by 15 fs laser pulse. As presented in Fig.~ \ref{fig2}(a) H17 is generated in Xe in the cut-off regime {\it only} for the laser frequencies higher than approximately $0.95 \omega_{Ti:Sapp}$. The intensities providing similar ionization for helium are higher than the ones shown in the Fig.~\ref{fig2}(b), so in this case the generation of H91 can take place in both regimes.  

\begin{figure}
\centerline{\includegraphics[width=0.9\columnwidth]{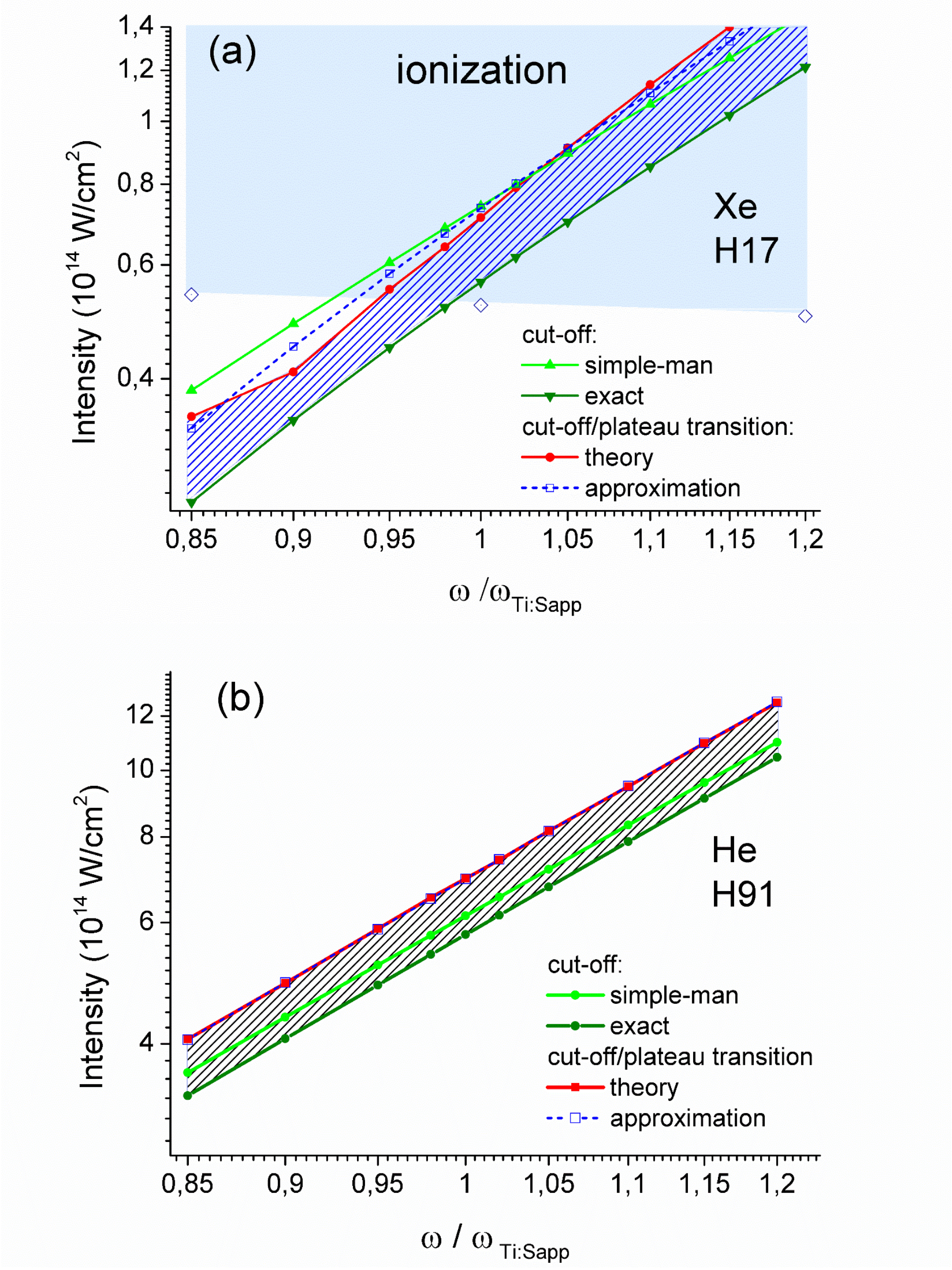}}
\caption{(color online) The intensities corresponding to the cut-off/plateau transition obtained theoretically (red line) and  approximately (blue dashed curve) for different fundamental frequencies. The simple-man and exact cut-off intensities (see text) are shown with light and dark green lines, respectively. The results are presented for (a) H17 generated by Xe atom and (b) H91 generated by He atom. The blue area shows the region with ionization more than 10\%.}
\label{fig2}
\end{figure}  

The second part of our study deals with the dephasing between the successive harmonics. The phase-locking of the harmonics makes possible the attosecond pulse generation. The cut-off region is especially important for the so-called amplitude gating method of an isolated attosecond pulse generation: the cut-off xuv is generated by a few-cycle laser pulse during one half-cycle only in the maximum of the laser pulse for the certain carrier envelope phase~\cite{amp_gating_christov,amp_cep_plat,amp_cep_paulus,amp_cep_exp,amp_gate_exp}. 

\begin{figure}
\centering{\includegraphics [width=0.8\columnwidth] {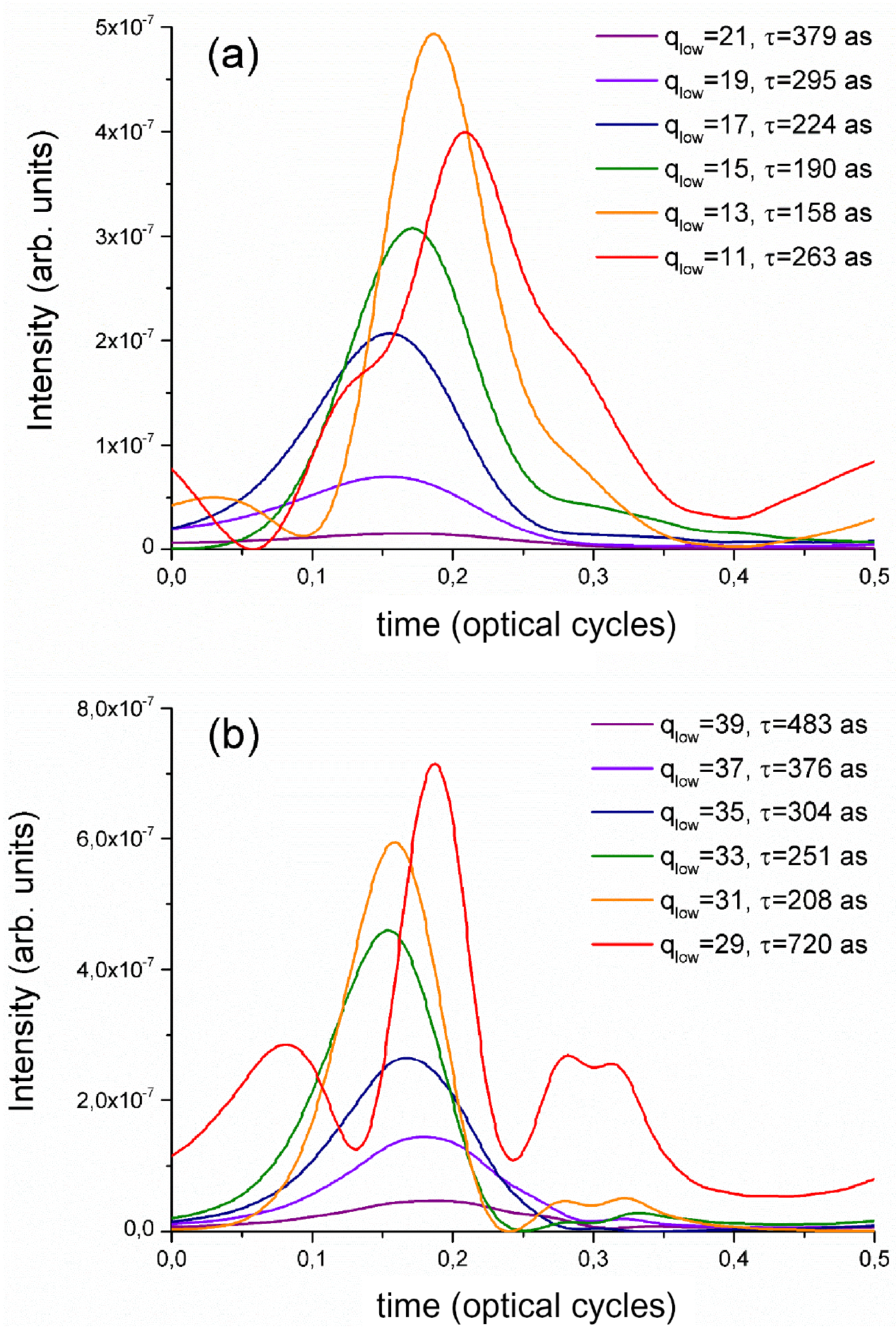}}
\caption{Calculated attosecond pulses obtained using all harmonics higher than $q_{low}$ shown in the legends. (a) Laser frequency is $1.5\omega_{Ti:Sapp}$ and its intensity 4.6 $\cdot$ 10$^{14}$ W/cm$^2$ corresponding to cut-off at 33$^{th}$ harmonic. (b) Laser frequency is $\omega_{Ti:Sapp}$ and its intensity 2.4 $\cdot$ 10$^{14}$ W/cm$^2$ corresponding to cut-off at 39$^{th}$ harmonic. }
\label{fig_a_p} 
\end{figure}

Using the numerical 3D TDSE solution for a model argon atom~\cite{Strelkov_Becker} we investigate the duration of the attosecond pulse from cut-off harmonics as a function of its spectral width and the fundamental frequency. In Fig.~\ref{fig_a_p} we present the attosecond pulses generated using different number of harmonics. We start form the case when the attosecond pulse is obtained using all harmonics above the cut-off one including it and then we add the adjacent lower harmonics one by one.  In Fig.~\ref{fig_a_p} one can see that initially the increasing number of harmonics leads to the decrease of the attosecond pulse duration. This fact is a result of the equal dephasing between successive harmonics resulting in the same emission time $t_e$ ($t_e=\frac{\partial \varphi}{\partial \omega}=\frac{\Delta \varphi}{2 \omega}$  where $\Delta \varphi$ is the dephasing (the spectral phase difference) between two consecutive harmonics \cite{mairesse_sci,emis_time}), thus the decrease of the attopulse duration is a manifestation of the uncertainty principle. However, adding lower harmonics with emission time differing from the one of the cut-off harmonics leads to the increase of the attopulse duration (see red curves in Fig.~\ref{fig_a_p}) and generation of two or more attosecond pulses. These pulses can be attributed to the short and long trajectory contributions in the plateau region~\cite{emis_time}. Below we denote $q_{low}$ corresponding to the shortest attosecond pulse duration as $q_{low}^{opt}$ and introduce the parameter $\beta=\frac{q_{cut-off}- q_{low}^{opt}+1}{q_{cut-off}}$ characterizing the ratio of the number of harmonics minimizing the attopulse duration to the total number of generated harmonics. For the conditions of Fig.~\ref{fig_a_p}(a) we find $\beta=0.43$ and for the ones of Fig.~\ref{fig_a_p}(b) $\beta=0.23$.  

\begin{figure}
\centering
\includegraphics [width=0.8\columnwidth] {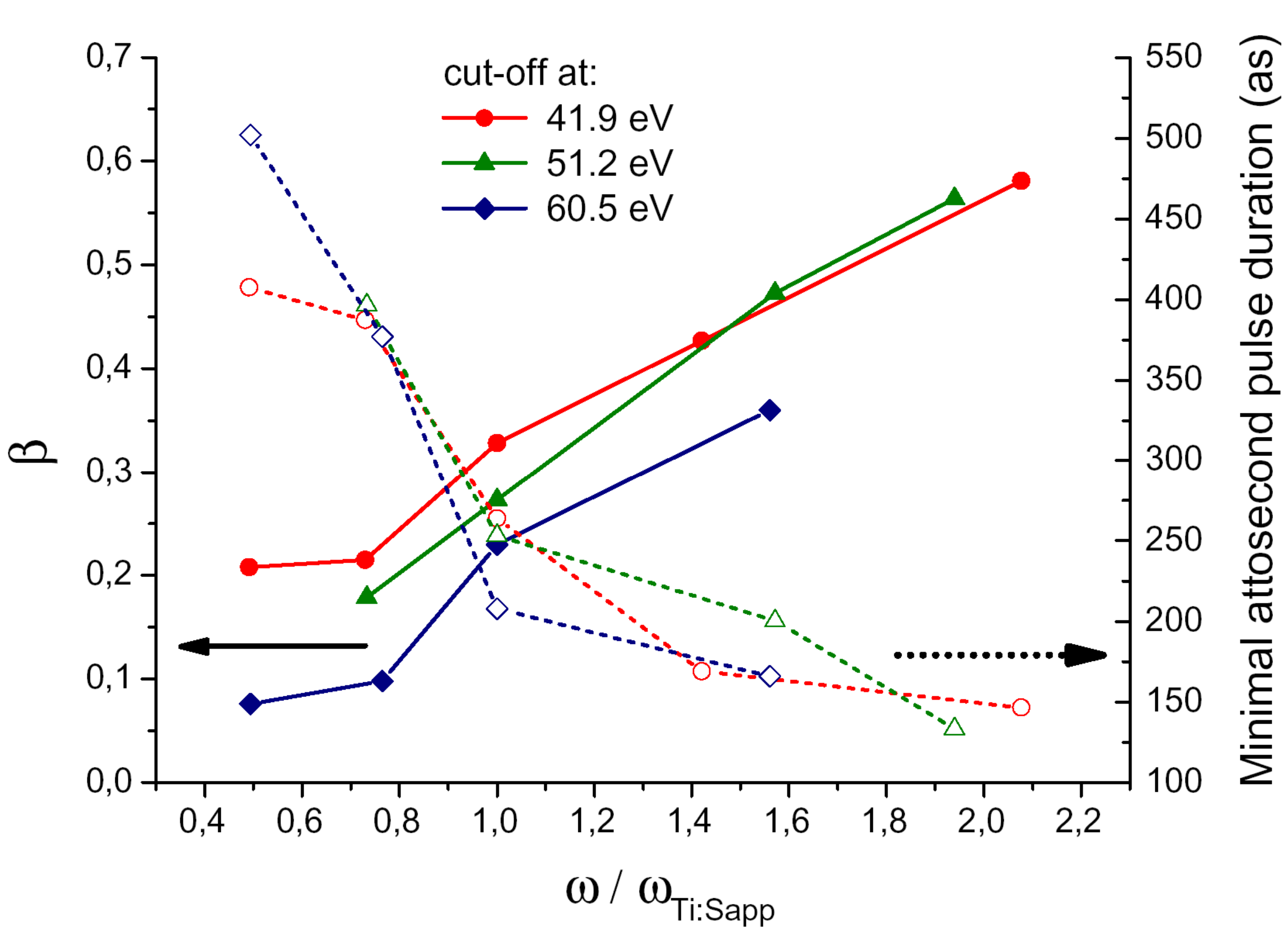}
\caption{(color online) Parameter $\beta$ (solid lines) and the single attosecond pulse duration (dashed lines) for certain values of $\beta$ for different laser intensities corresponding to 41.9 eV (red), 51.2 eV (green) and 60.5 eV (blue) cut-off harmonic energies.}
\label{fig_t_e} 
\end{figure}

In Fig.~\ref{fig_t_e} the values for $\beta$ are shown for different intensities and frequencies of the laser field. These are chosen as follows: first we calculate the spectrum for the laser field with wavelength of 800 nm and intensities corresponding to the simple-man cut-off frequency at 27H, 33H and 39H. The energies of the cut-off harmonics are 41.9 eV, 51.2 eV and 60.5 eV, respectively. Furthermore, we consider other laser frequencies and intensities so that the cut-off takes place at these energies.

One can see that the parameter $\beta$ (solid lines in Fig.~\ref{fig_t_e}) increases with increasing fundamental frequency while the corresponding attosecond pulse shortens. The shortest attopulse duration obtained in our calculations is less than 150 as. Moreover, as the generation conditions get closer to the tunneling regime the $\beta$ parameter decreases. At the same time, the minimal attosecond pulse duration for the given frequency almost doesn't depend on the fundamental intensity. The shortest attosecond pulse duration is 0.08 -- 0.1 of the laser cycle for different intensities and frequencies of the fundamental. For the frequency of the Ti:Sapp laser this agrees very well with the experimentally found duration of 250~as~\cite{amp_gate_exp}. 

Thus, in this Letter we find that the cut-off regime can be defined as one having regular linear dependence of the harmonic phase on the fundamental intensity. The phase coefficient in the cut-off regime is well approximated with equation~(\ref{alpha_cut-off}). The phase coefficient grows as the cube of the fundamental wavelength, therefore this dependence becomes very important for the HHG by mid-infrared fields. The value of the phase coefficient is much higher than the one for the short trajectory in the plateau regime. We show that HHG takes place in the cut-off regime for a {\it range} of intensity and the microscopic response amplitude achieves significant magnitude (in some cases it achieves its maximum) within this range. Moreover, for rather high harmonics the generation occurs mainly within this range because the medium ionization practically limits the fundamental intensity for which the HHG takes place.  

Change of the harmonic phase-locking when HHG evolves from the cut-off to the plateau regime determines the optimal bandwidth of the spectral region which should be used for the attosecond pulse production via amplitude gating technique. We find that in argon the minimal pulse duration which can be obtained with this technique without using dispersion elements is approximately 0.08 -- 0.1 of the laser cycle for different intensities and frequencies of the fundamental. Using filters with proper dispersion~\cite{Lopez-Martens} or chirped multilayer mirrors~\cite{Goulielmakis} is necessary to obtain shorter attosecond pulses.  

This study was supported by RFBR (grants N 15-32-50361 and N 14-02-00878).


%% file: appendix.txt
\section*{Appendix}

We solve numerically the one-electron 3D TDSE for a model Xe atom in the laser field. The numerical method is described in~\cite{ref1}. The soft Coulomb potential providing the ionization energy equal to the Xe ionization energy is used.   We reconstruct the $\alpha$ values from the TDSE solution as discussed below.

{\it Method 1} is a development of the method suggested in Refs.~\cite{ref2,ref3} where $\alpha$ was reconstructed from the spectra emitted under different peak intensities of the Gaussian laser pulse. However, we use very specific temporal envelope of the laser intensity: the intensity rapidly increases at the leading edge of the pulse during time $\tau_f$, then it slowly {\it linearly} grows during time $\tau$, and finally it decreases during time $\tau_f$. The intensity is given as:

\begin{equation}
I(t)=I_0 f(t) \left(1+\gamma \frac{\omega t}{2 \pi} \right)
\label{int1}
\end{equation} 
where
\begin{equation}
f(t)=\left\{
\begin{array}{rcl}
\\-\tau/2-\tau_f<t<-\tau/2 & , &\sin^2\left( \frac{t+\tau/2+\tau_f}{\tau_f} \frac{\pi}{2}\right)
\\-\tau/2<t<\tau/2 & , &1
\\ \tau/2<t<\tau/2+ \tau_f & , &\sin^2\left( \frac{t-[\tau/2+\tau_f]}{\tau_f} \frac{\pi}{2}\right)
\end{array}
\right.
\label{flat-top}
\end{equation} 

The harmonics are emitted mainly in the region of the linear growth of the intensity.  The linear dependence of the harmonic phase on the laser intensity is transmitted in the linear dependence of the phase on time, thus in the harmonic frequency shift. Presence of several contributions to HHG leads to the line splitting, as it is shown in Fig.~\ref{Fig2_app}. The value of $\alpha$ for every contribution can be reconstructed from the frequency shift of every peak (as well as the weight of the contribution can be reconstructed from the weight of the peak). Namely,  the spectral component with the frequency shift $\delta \omega$ corresponds to
\begin{equation}
\alpha=2 \pi \frac{\delta \omega}{\omega} \frac{1}{\gamma I_0}
\end{equation}

\begin{figure}  
\centering
\includegraphics [width=0.7\columnwidth] {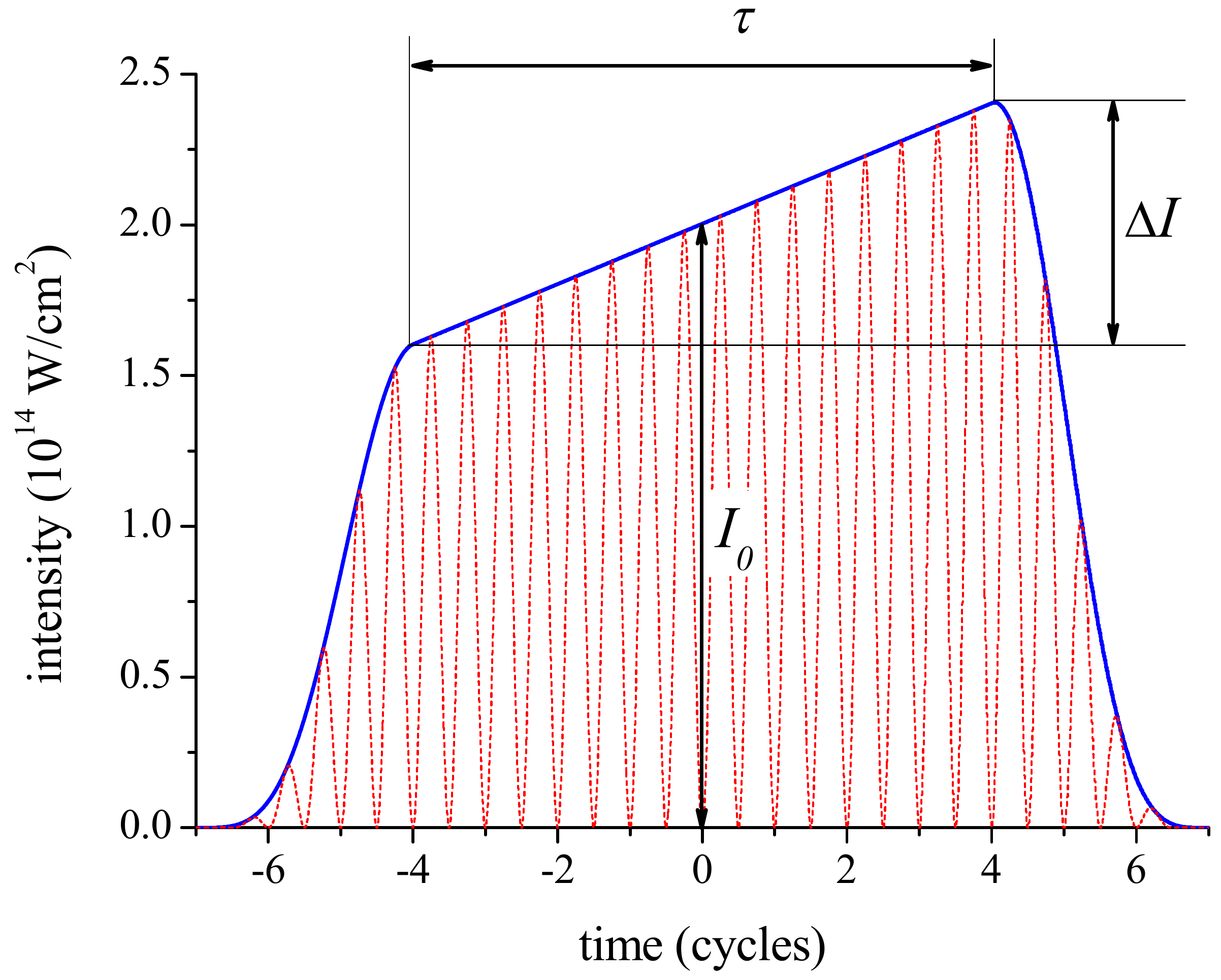}
\caption{(color online) 
The laser pulse intensity (solid blue line) given by the equation~(\ref{int1}).  After $\tau_f=3$ cycles turning on, the intensity increases linearly during $\tau=8$ cycles, and then turns off during 3 cycles. Dotted red line presents the square of the field.}
\label{Fig1_app}
\end{figure}

\begin{figure}  
\centering
\includegraphics [width=0.7\columnwidth] {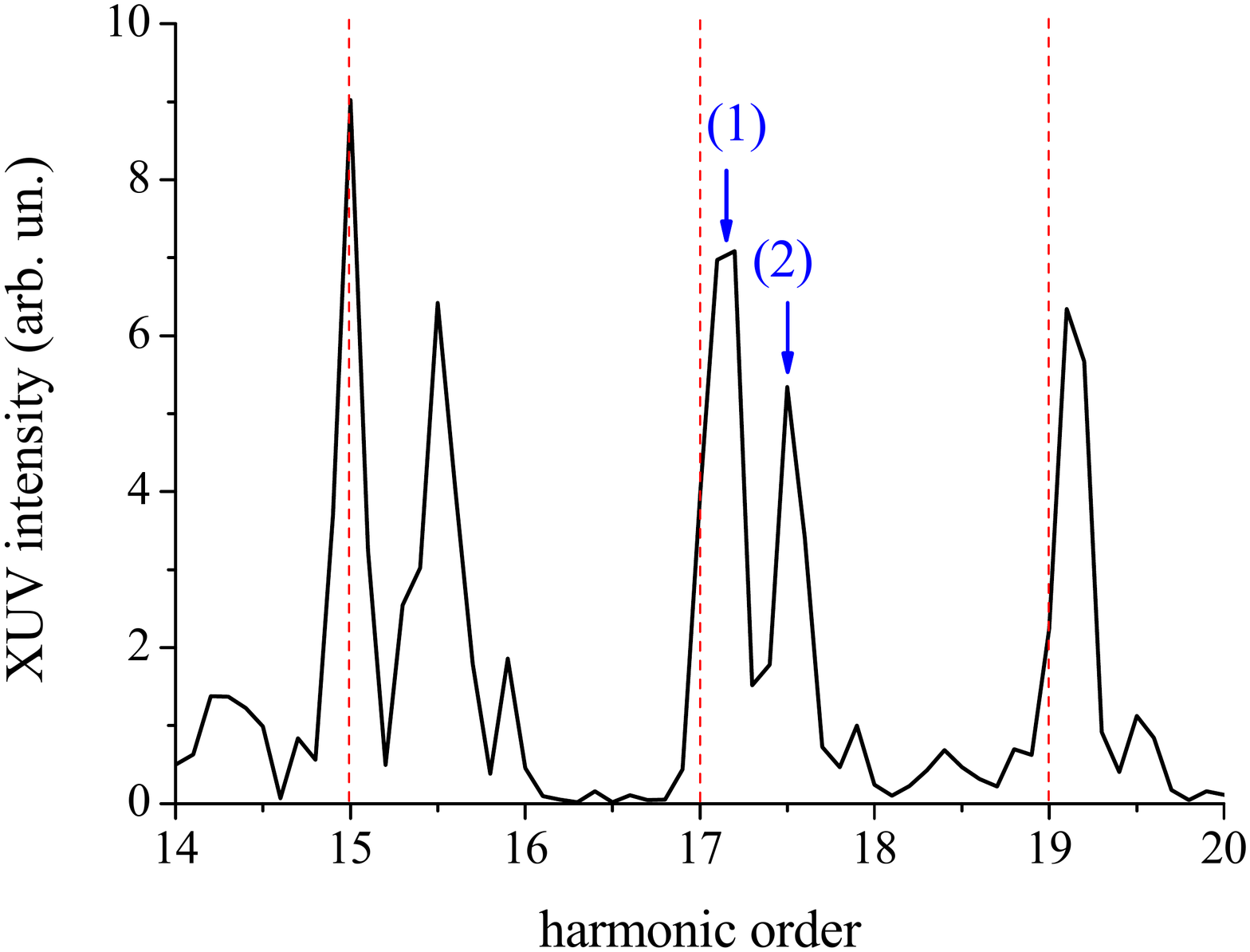}
\caption{(color online) 
Harmonic spectrum found via numerical 3D TDSE solution for the model Xe atom. The laser wavelength is 800 nm and the laser intensity envelope is shown in Fig.~\ref{Fig1_app}. Red dashed lines show exact harmonic frequencies. One can see the shortest (1), next-to-the-shortest (2), and other trajectories contributions.}
\label{Fig2_app}
\end{figure}

The accuracy of the method is limited with the following: using too short $\tau$ leads to the wide spectral peaks (due to the uncertainty principle), and using too long $\tau$ provides full ionization and high variation of the intensity within the pulse $\Delta I$, so it is unclear which laser intensity corresponds to the found $\alpha$. In general, this method is effective for the contributions with high $\alpha$: even slow intensity growth leads to the pronounced shift of the spectral peak.  

\begin{figure*}
\centering
\includegraphics [width=1.4\columnwidth] {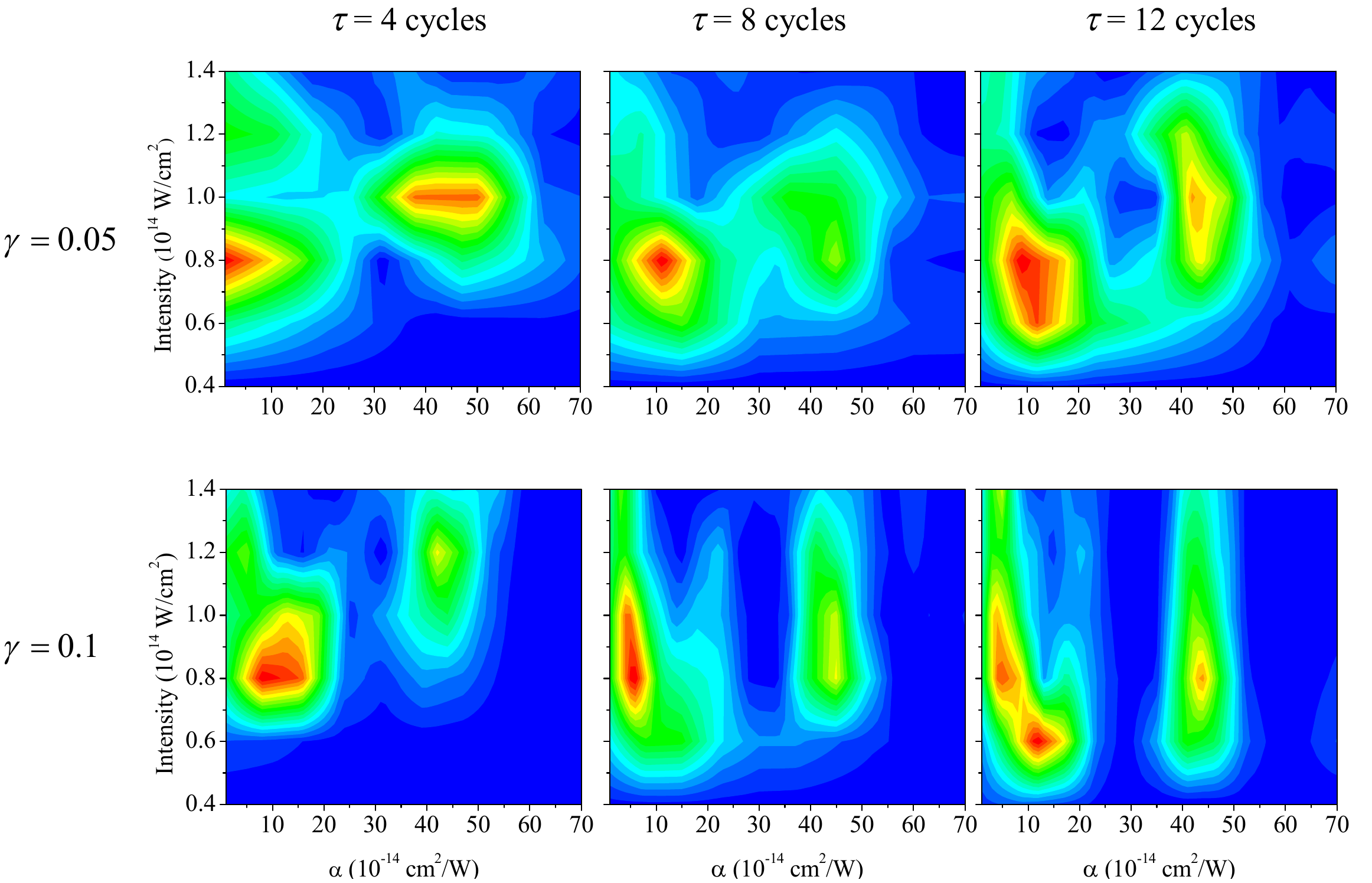}
\caption{(color online) 
$\alpha$ maps calculated for different $\tau $ and $\gamma$ for H17 in Xe, the laser wavelength is 800 nm.}
\label{Fig3_app}
\end{figure*}

In Fig.~\ref{Fig3_app} we compare $\alpha$ maps found for different $\tau $ and $\gamma$. One can see that the features of the maps are rather sensitive to these parameters. Certainly, this is not a drawback of the method but a manifestation of the uncertainty principle. The main properties are similar for all the presented maps: when the harmonic is generated in the cut-off regime it is rather intense and the $\alpha$ value is about ten. For higher intensities there are several contributions with different $\alpha$ values. In particular, in the lower right panel (having the best resolution in terms of $\alpha$ and, thus the worst resolution in terms of intensity) we can clearly see the two contributions corresponding to the short ($\alpha \approx 3$) and long  ($\alpha \approx 20$) trajectories for the electronic excursion time less then one cycle and (surprisingly intense) contribution from the trajectory with longer excursion time  ($\alpha \approx 45$). 

\begin{figure}
\centering
\includegraphics [width=0.7\columnwidth] {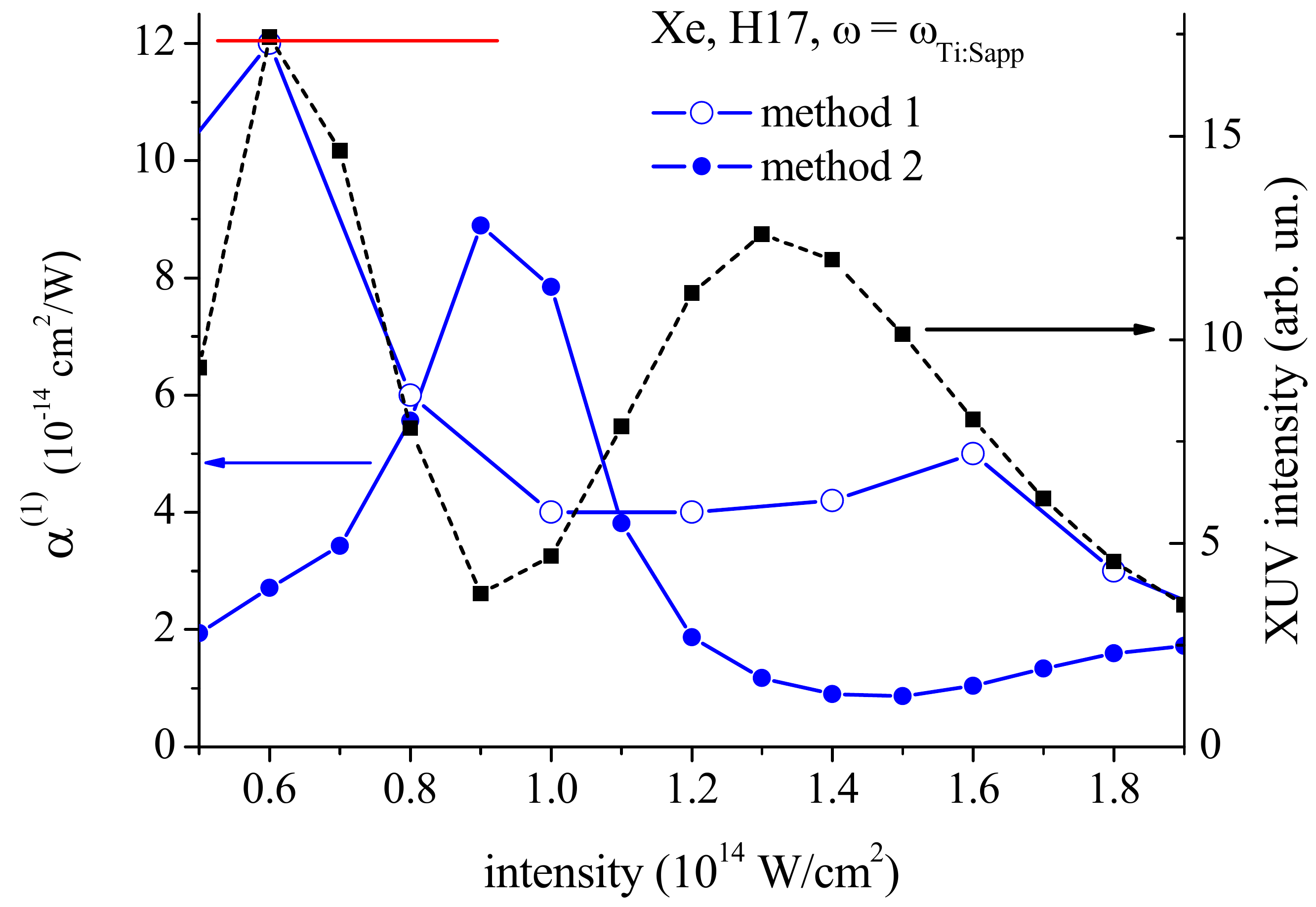}
\caption{(color online) 
 $\alpha^{(1)}$ for the shortest trajectory contribution (solid blue) and the intensity (dotted black) of this contribution. The thick red line shows $\alpha$ for the cut-off harmonics given by equation (5), see the main text of the Letter.
 Results of method 1 (open circles) and method 2 (solid circles) calculated for H17 generated in Xe for the laser wavelength 800 nm.}
\label{Fig4_app_a}
\end{figure}

\begin{figure}
\centering
\includegraphics [width=0.7\columnwidth] {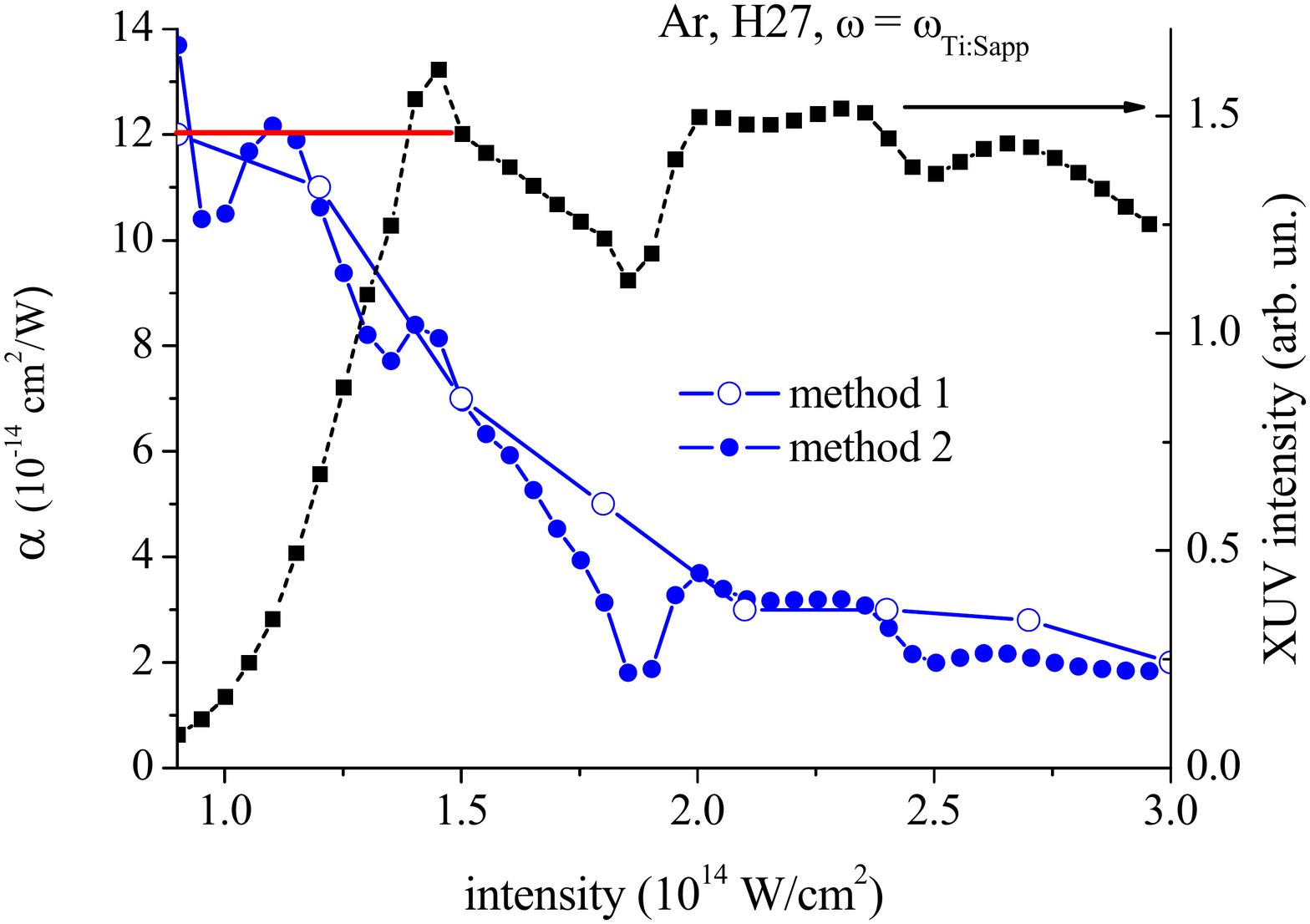}
\caption{(color online) 
The same as Fig.~\ref{Fig4_app_a} for H27 generated in Ar by 800 nm laser field.}
\label{Fig4_app_b}
\end{figure}

\begin{figure}
\centering
\includegraphics [width=0.7\columnwidth] {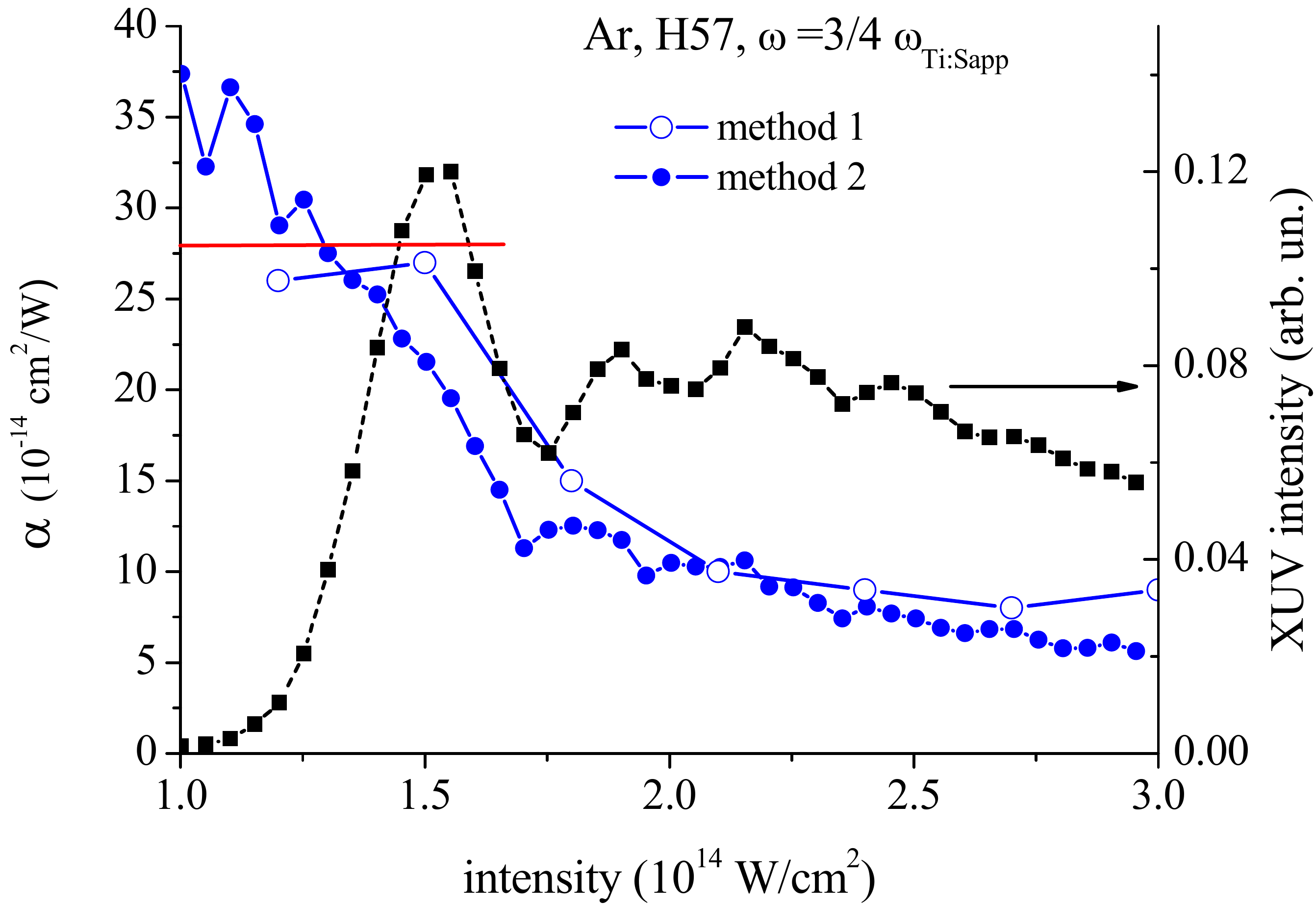}
\caption{(color online) 
The same as Fig.~\ref{Fig4_app_a} for H57 generated in Ar by 1067 nm laser field.}
\label{Fig4_app_c}
\end{figure}

{\it Method 2} is based on the TDSE numerical solution in which all trajectories contributions except the shortest one are artificially suppressed. The technique of this suppression was described in Ref.~\cite{ref4}. In this method specific shape of the the laser pulse is not important. We use the intensity given by:
\begin{equation}
I(t)=I_0 f(t) 
\label{int2}
\end{equation} 
where $f(t)$ is given by the equation~(\ref{flat-top}), $\tau_f$ is 3 laser cycles and $\tau$ is 4 cycles.

From the harmonic phase $\varphi^{(1)}(I_0)$ (the index $(1)$ shows that only the shortest trajectory contribution is taken into account in this method) calculated for different $I_0$  we then find $\alpha^{(1)}=-\partial \varphi^{(1)}/ \partial I_0$, see Figs.~\ref{Fig4_app_a}-\ref{Fig4_app_c}. The harmonic intensity found in these calculation gives the intensity of the shortest trajectory contribution as the function of the laser intensity. 

{\it Comparison} of the methods is presented in  Figs.~\ref{Fig4_app_a}-\ref{Fig4_app_c}. To find $\alpha^{(1)}$ with method 1 we choose the peak corresponding to the lowest $\alpha$ value for every intensity. The results of both methods are very close to each other in the tunneling regime (Figs.~\ref{Fig4_app_b}-\ref{Fig4_app_c}) and are less close for conditions intermediate between tunnel and multiphoton regimes (Fig.~\ref{Fig4_app_a}). The method 2 is less reliable in these conditions because the quantum trajectories separation used in this method is hardly applicable. However, in all 3 cases for the harmonic generated in the cut-off regime method 1 gives $\alpha$ value which is close to the theoretical estimation given by equation~(5), see the main text of the Letter.

The xuv intensity emitted in the cut-off regime is comparable (Fig.~\ref{Fig4_app_b})  or even higher  (Figs.~\ref{Fig3_app},~\ref{Fig4_app_a},~\ref{Fig4_app_c}) than in the plateau regime. This makes the studies of the cut-off regime rather important from the practical viewpoint.

Note, that  the sensitivity of the harmonic phase on the laser intensity influences many HHG features like phase-matching, divergence, etc. The rapid increase of the $\alpha$ value with the laser wavelength makes this sensitivity even more important for the HHG in few-micron laser fields which are recently actively used for generation of very high harmonic orders, see Refs~\cite{Sheehy1999,Shan2001} and others.    
